\documentstyle[preprint,revtex]{aps}
\tightenlines
\begin{document}
\draft

\preprint{SLAC--PUB--5992}
\medskip
\preprint{November 1992}
\medskip
\preprint{T/E}

\begin{title}
Two-loop corrections to the Isgur-Wise function\\
in QCD sum rules\thanks{Supported by the Department of Energy under
contract DE-AC03-76SF00515.}
\end{title}

\author{Matthias Neubert}
\begin{instit}
Stanford Linear Accelerator Center\\
Stanford University, Stanford, California 94309
\end{instit}
%\receipt{November 1992}

\begin{abstract}
We complete the QCD sum rule analysis of the Isgur Wise form factor
$\xi(v\cdot v')$ at next-to-leading order in renormalization-group
improved perturbation theory. To this end, the exact result for the
two-loop corrections to the perturbative contribution is derived using
the heavy quark effective theory. Several techniques for the evaluation
of two-loop integrals involving two different types of heavy quark
propagators are discussed in detail, among them the methods of
integration by parts and differential equations. The order-$\alpha_s$
corrections to the Isgur-Wise function turn out to be small and well
under control. At large recoil, they tend to decrease the form factor
by $5-10\%$.
\end{abstract}
%\pacs{11.50.Li,12.38.Bx,12.38.Cy,11.30.Ly}

\centerline{(Submitted to Physical Review D)}
\newpage
\narrowtext

\section{Introduction}

In the limit of infinite heavy quark masses, the weak decay form
factors describing semileptonic transitions between any two
ground-state (pseudoscalar or vector) heavy mesons, $M(v)\to M'(v')\,
\ell\,\nu$, are described by a universal form factor $\xi(y)$. This
so-called Isgur-Wise function depends on the velocity transfer
$y=v\cdot v'$ and is normalized at zero recoil, $\xi(1)=1$, where the
initial and final meson have the same velocity \cite{Isgu}. These
remarkable results follow from an implicit spin-flavor symmetry, which
QCD reveals for heavy quarks although it is not explicit from its
Lagrangian \cite{Volo}. An expansion about this symmetry limit is
afforded by the construction of the heavy quark effective theory (HQET)
\cite{Geor,Mann,Falk}. In the infinite mass limit, its effective
Lagrangian is explicitly invariant under spin-flavor symmetry
transformations. HQET thus provides a convenient framework in which to
analyze the properties of hadrons containing a heavy quark. In
particular, it allows a systematic expansion of weak decay form factors
in powers of $1/m_Q$ \cite{Falk,Luke,AMM}.

To leading order in this expansion one recovers the Isgur-Wise limit,
in which a large set of otherwise unrelated form factors reduces to the
Isgur-Wise function. This function describes the dynamical properties
of the cloud of light quarks and gluons surrounding the static heavy
quarks. Being a hadronic form factor, it can only be investigated
using nonperturbative methods. One such method is provided by QCD sum
rules \cite{SVZ}, which were originally developed for light quark
systems and have yielded many nice results which are competitive with
lattice computations. Recently, several authors have used QCD sum rules
to calculate hadronic matrix elements in HQET
\cite{Buch,Rady,BGSR,SR,BBBD,Sublea,Blok,Baier,chi2,Patricia}. The more
refined of these analyses included radiative corrections to both the
perturbative and nonperturbative contributions. For the Isgur-Wise
function, however, the radiative corrections to the perturbative part
of the sum rule were only incorporated in leading logarithmic
approximation \cite{SR}; the complete two-loop corrections to the
triangle quark loop were never calculated for the case of heavy
quarks. On the other hand, from the well-studied case of pseudoscalar
decay constants it is known that next-to-leading logarithmic
corrections can be quite substantial and should in principle be taken
into account \cite{BGSR,SR,BBBD}.

With the advance of HQET, considerable progress has been made in the
calculation of radiative corrections. In Ref.~\cite{chi2}, for the
first time an exact two-loop result was obtained for a heavy meson form
factor, in this case for one of the universal functions that appear at
order $1/m_Q$ in the heavy quark expansion. The authors of
Ref.~\cite{Patricia} developed a general method to compute the first
two terms in an expansion of a two-loop diagram in HQET as a power
series in $(y-1)$, and applied their technique to obtain an expansion
of the two-loop corrections to the perturbative part of the Isgur-Wise
function close to zero recoil.

In this paper we derive the exact result for the two-loop corrections
to $\xi(y)$. To this end, we develop several techniques to evaluate
two-loop integrals in HQET involving two heavy quarks with different
velocities ($v$ and $v'$) and different residual momenta ($k$ and
$k'$), among them the method of integration by parts
\cite{IbP1,IbP2,IbP3} and the use of differential equations \cite{DEq}.
We also introduce an integral representation for a general two-loop
diagram which is particularly convenient for QCD sum rule calculations.
These techniques are rather general and can be readily applied to other
cases. In Sec.~\ref{sec:2} we briefly review the sum rule analysis of
the Isgur Wise function. The calculation of the two-loop perturbative
corrections to $\xi(y)$ are described in Sec.~\ref{sec:3}. We discuss
in detail the contribution of each individual diagram, so that the
interested reader can follow the analysis step by step. After
renormalization, we compare our exact result with the expansion around
zero recoil given in Ref.~\cite{Patricia} and find agreement.
Sec.~\ref{sec:4} deals with the renormalization-group improvement and
the numerical analysis of the sum rule. We find that the effects of
radiative corrections to the Isgur-Wise functions are moderate and well
under control. Sec.~\ref{sec:5} contains the conclusions.

\section{Sum rule for the Isgur-Wise function}
\label{sec:2}

The derivation of the QCD sum rule for the Isgur-Wise function has been
dealt with at length in Refs.~\cite{Rady,SR,Sublea}, to which we refer
the interested reader for details. Here we restrict ourselves to a
brief review for the purpose of introducing the necessary notations and
recalling the main ideas of the method. One studies the analytic
properties of the three-current correlator
\begin{eqnarray}\label{correl}
   &&\int\!{\rm d}x\,{\rm d}z\,e^{i(k'\cdot x - k\cdot z)}\,
    \langle 0\,|\,T\Big\{
    \big[\, \bar q\,\overline{\Gamma}_{M'} h' \,\big]_x,
    \big[\, \bar h'\,\Gamma\,h \,\big]_0,
    \big[\, \bar h\,\Gamma_M q \,\big]_z \Big\} |\,0\,\rangle
    \nonumber\\
   &&\equiv \Xi(\omega,\omega',v\cdot v')\,
    {\rm Tr}\big\{ \overline{\cal{P}}'\Gamma\,{\cal{P}} \big\} \,,
\end{eqnarray}
where $h$ and $h'$ describe heavy quarks with velocities $v$ and $v'$
in the effective theory. These quarks have ``residual'' momenta $k$ and
$k'$, which are related to the total external momenta by $P=m_Q v+k$
and $P'=m_{Q'} v'+k'$, where $m_Q$ and $m_{Q'}$ are the heavy quark
masses. Depending on the choice $\Gamma_M = -\gamma_5$ or $\Gamma_M =
\gamma_\mu - v_\mu$, the heavy-light currents interpolate pseudoscalar
or vector mesons, respectively. The Dirac structure $\Gamma$ of the
heavy-heavy current is arbitrary. Usually, however, this is a
flavor-changing weak current, in which case $\Gamma=\gamma_\mu
(1-\gamma_5)$. The Dirac structure of the correlator is entirely
contained in the trace over ``spin wave functions''
\begin{equation}
   {\cal{P}}={1+\rlap/v\over 2}\,\Gamma_M ~,~~
   \overline{\cal{P}}' = \overline{\Gamma}_{M'}\,
   {1+\rlap/v'\over 2} \,,
\end{equation}
which act as projection operators:
\begin{eqnarray}
   \rlap/v\,{\cal{P}} &=& {\cal{P}} = - {\cal{P}}\,\rlap/v \,,
    \nonumber\\
   -\rlap/v'\,\overline{\cal{P}}' &=& \overline{\cal{P}}'
    = \overline{\cal{P}}'\rlap/v' \,.
\end{eqnarray}

The coefficient function $\Xi(\omega,\omega',v\cdot v')$ in
(\ref{correl}) is analytic in the ``off-shell energies'' $\omega=
2 v\cdot k$ and $\omega'=2 v'\cdot k'$, with discontinuities for
positive values of these variables. In particular, it receives a
double-pole contribution from the ground-state mesons $M$ and $M'$
associated with the heavy-light currents. This pole is located at
$\omega=\omega'= 2\bar\Lambda$, where $\bar\Lambda = m_M-m_Q =
m_{M'}-m_{Q'}$. The residue is proportional to the Isgur-Wise function.
It follows that \cite{SR}
\begin{equation}\label{pole}
   \Xi_{\rm pole}(\omega,\omega',y) = {\xi(y,\mu)\,F^2(\mu)\over
   (\omega-2\bar\Lambda+i\epsilon) (\omega'-2\bar\Lambda+i\epsilon)}
   \,,
\end{equation}
where $y=v\cdot v'$, and $F$ corresponds to the scaled meson decay
constant in the effective theory ($F\sim f_M\sqrt{m_M}$). Note that
both $F$ and the Isgur-Wise function are defined in terms of matrix
elements in the effective theory and are therefore scale-dependent
quantities.

In the deep Euclidean region, the correlator can be calculated
perturbatively by using the Feynman rules of the heavy quark effective
theory \cite{Falk}. The idea of QCD sum rules is that, at the
transition from the perturbative to the nonperturbative regime,
confinement effects can be accounted for by including the leading power
corrections in the operator product expansion of the three-point
function. They are proportional to vacuum expectation values of local
quark-gluon operators, the so-called condensates \cite{SVZ}. One then
writes the theoretical expression for the correlator in terms of a
double dispersion integral,
\begin{equation}
   \Xi_{\rm th}(\omega,\omega',y) = \int\!{\rm d}\nu\,{\rm d}\nu'\,
   {\rho_{\rm th}(\nu,\nu',y)\over(\nu-\omega-i\epsilon)
   (\nu'-\omega'-i\epsilon)} + {\rm subtractions} ,
\end{equation}
and performs a Borel transformation in $\omega$ and $\omega'$ (see
Appendix~\ref{app:1} for the definition of the Borel operator). This
yields an exponential damping factor in the dispersion integral and
eliminates possible subtraction terms. Because of the flavor symmetry
it is natural to set the associated Borel parameters equal:
$\tau=\tau'\equiv 2 T$. Following Refs.~\cite{SR,Blok}, one then
introduces new variables $\omega_\pm=\case{1}/{2}(\nu\pm\nu')$,
performs the integral over $\omega_-$, and employs quark-hadron duality
to equate the integral over $\omega_+$ up to a threshold $\omega_0$ to
the Borel transform of the pole contribution in (\ref{pole}). This
yields the Borel sum rule
\begin{equation}\label{sum}
   \xi(y,\mu)\,F^2(\mu)\,e^{-2\bar\Lambda/T}
   = \int\limits_0^{\omega_0}\!{\rm d}\omega_+\,e^{-\omega_+/T}\,
   \widetilde{\rho}_{\rm\,th}(\omega_+,y) \equiv K(T,\omega_0,y) \,.
\end{equation}
The effective spectral density $\widetilde{\rho}_{\rm\,th}$ arises
after integration of the double spectral density over $\omega_-$.

To lowest order in perturbation theory, the theoretical expression for
the right-hand side of the sum rule is given by \cite{Buch,Rady,SR}
\begin{eqnarray}
   K(T,\omega_0,y) &=& {3\over 8\pi^2}\bigg({2\over y+1}\bigg)^2
    \int\limits_0^{\omega_0}\!{\rm d}\omega_+\,\omega_+^2\,
    e^{-\omega_+/T} - \langle\bar q q\rangle \nonumber\\
   &&+ \bigg({y-1\over y+1}\bigg)
    {\langle\alpha_s GG\rangle\over 48\pi T} + {(2y+1)\over 3}\,
    {m_0^2\,\langle\bar q q\rangle\over 4 T^2} \,.
\end{eqnarray}
We have included the leading nonperturbative contributions in the
operator product expansion, which are proportional to the quark
condensate (dimension $d=3$), the gluon condensate ($d=4$), and the
mixed quark-gluon condensate ($d=5$). In the numerical analysis in
Sec.~\ref{sec:4} we will use the standard values (at $\mu=1$ GeV)
\begin{eqnarray}\label{cond}
   \langle\bar q q\rangle &=& -(0.23\,{\rm GeV})^3 \,, \nonumber\\
   \langle\alpha_s GG\rangle &=& 0.04\,{\rm GeV^4} \,, \nonumber\\
   \langle g_s\bar q\sigma_{\mu\nu}G^{\mu\nu}q\rangle &=&
    m_0^2\,\langle\bar q q\rangle ~,~~ m_0^2=0.8\,{\rm GeV^2} \,.
\end{eqnarray}

At zero recoil, a Ward identity relates the three-point function
(\ref{correl}) to the correlator of two heavy-light currents, from
which one derives the sum rule for the parameter $F$. It allows one to
replace the product $F^2 e^{-2\bar\Lambda/T}$ in (\ref{sum}) by
$K(T,\omega_0,1)$. Then the final form of the sum rule for the
Isgur-Wise function explicitly reveals its normalization at zero
recoil:
\begin{equation}
   \xi(y,\mu) = {K(T,\omega_0,y)\over K(T,\omega_0,1)} \,.
\end{equation}

In the following section we will derive the complete expression for the
perturbative corrections to the function $K(T,\omega_0,y)$ arising at
order $\alpha_s$. The one-loop corrections to the quark condensate were
calculated in Ref.~\cite{SR}. The mixed and gluon condensate are
already of order $g_s$ or $g_s^2$, and consequently one does not have
to include radiative corrections to these terms at order $\alpha_s$.
What is missing are thus the order-$\alpha_s$ corrections to the
perturbative contribution. There are restrictive constraints on the
result of this two-loop calculation. The normalization of the
Isgur-Wise function requires that, at zero recoil, one must recover the
expression for the two-loop corrections to the perturbative part of the
sum rule for $F$. This implies \cite{BGSR,SR,BBBD}
\begin{equation}\label{Kpert1}
   K_{\rm pert}(T,\omega_0,1) = {3\over 8\pi^2}\,
   \int\limits_0^{\omega_0}\!{\rm d}\omega_+\,\omega_+^2\,
   e^{-\omega_+/T}\,\bigg\{1 + {\alpha_s\over\pi} \bigg[
   2\ln{\mu\over\omega_+} + {4\pi^2\over 9} + {17\over 3} \bigg]
   \bigg\} \,.
\end{equation}
Furthermore, the two-loop calculation must reproduce the known
anomalous dimension of both $F(\mu)$ and $\xi(y,\mu)$
\cite{Volo,Poli,Falk}. We will see at the end of Sec.~\ref{sec:3} how
these constraints are fulfilled.

\section{Two-loop calculation}
\label{sec:3}

The two-loop corrections to the perturbative contribution to the
three-current correlator (\ref{correl}) are shown in Fig.~\ref{fig:1}.
We shall analyze these diagrams separately below. Throughout the
calculation we use Feynman gauge. For practical purposes, it is useful
to realize that the dependence of the perturbative spectral density
$\widetilde{\rho}_ {\rm pert}(\omega_+,y)$ on $\omega_+$ is known on
dimensional grounds:
\begin{eqnarray}\label{wdep}
   \widetilde{\rho}_{\rm pert}(\omega_+,y) &=& \omega_+^2
    \bigg[ \rho_1(y) + \rho_2(y)\,\ln{\mu\over\omega_+} \bigg] \,,
    \nonumber\\
   \Rightarrow\quad \int\limits_0^\infty\!{\rm d}\omega_+\,
   e^{-\omega_+/T}\,\widetilde{\rho}_{\rm pert}(\omega_+,y) &=&
    2 T^3 \bigg[ \rho_1(y) + \rho_2(y)\bigg(\ln{\mu\over T} +
    \gamma_E - {3\over 2}\bigg) \bigg] \,.
\end{eqnarray}
The coefficient functions $\rho_i(y)$ are independent of $\omega_+$.
It thus suffices to calculate directly the Borel transform of the
correlator, corresponding to the second line in this equation. The
spectral density can then be read off immediately. Below, we will
always denote Borel-transformed quantities by a ``hat''.

\subsection{\bf Gluons attached to heavy quark lines}

Consider the three diagrams $D_1$ to $D_3$ in Fig.~\ref{fig:1}. The
evaluation of the first graph gives
\begin{eqnarray}\label{D1}
   D_1 &=& 16 N_c C_F\,g_s^2\,y\,{\rm Tr}\big\{\,
    \overline{\cal{P}}'\Gamma\,{\cal{P}}\,\gamma^\alpha\big\}
    \nonumber\\
   &&\times \int\!{\rm d}\tilde s\,{\rm d}\tilde t\,
    {s_\alpha\over(\omega+2v\cdot s) (\omega+2v\cdot t)
    (\omega'+2v'\cdot s) (\omega'+2v'\cdot t)\,s^2\,(s-t)^2} \,,
\end{eqnarray}
where $C_F=(N_c^2-1)/2 N_c$, and ${\rm d}\tilde s\equiv (2\pi)^{-D}
{\rm d}^Ds$. The two-loop integral is most conveniently performed by
using a Fourier representation for the light quark and gluon
propagators, and an exponential integral representation for the heavy
quark propagators. A detailed description of this method, as well as
its application to the above integral, can be found in
Appendix~\ref{app:1}. After Borel transformation we find
\begin{equation}\label{D1res}
   \hat D_1 = -{4 y A\over(D-4)}\,\big[2(y+1)\big]^{D/2-2}
   G\big(0,0,\case{D}/{2}-1;y\big)
\end{equation}
where we have abbreviated
\begin{equation}\label{Adef}
   A = - {16 N_c C_F\,g_s^2\over(4\pi)^D}\,
   {(2T)^{2D-5}\over\big[2(y+1)\big]^{D-2}}\,
   \Gamma\big(\case{D}/{2}\big)\,\Gamma\big(\case{D}/{2}-1\big)\,
   {\rm Tr}\big\{\,\overline{\cal{P}}'\Gamma\,{\cal{P}}\,\big\} \,.
\end{equation}
Here and in the following we will often encounter integrals of the form
\begin{equation}
   G(a,b,c;y) = \int\limits_0^1\!{\rm d}u\,
   {(1-u)^a\,u^b\over\big(1+2yu+u^2\big)^c} \,,
\end{equation}
which are related to generalized hypergeometric functions of the
velocity transfer $y$.

The contribution of the second diagram is
\begin{eqnarray}
   D_2 &=& 16 N_c C_F\,g_s^2\,{\rm Tr}\big\{\,
    \overline{\cal{P}}'\Gamma\,{\cal{P}}\,\gamma^\alpha\big\}
    \nonumber\\
   &&\times \int\!{\rm d}\tilde s\,{\rm d}\tilde t\,
    {s_\alpha\over(\omega+2v\cdot s)^2 (\omega+2v\cdot t)
    (\omega'+2v'\cdot s)\,s^2\,(s-t)^2} \,.
\end{eqnarray}
Introducing a shifted loop momentum by $t'=t+s$, one can first perform
the integral over $t'$ and then carry out the integral over $s$ by
using the master equations (\ref{master1}) and (\ref{master2}) for
one-loop integrals in HQET (see Appendix~\ref{app:1}). After Borel
transformation, the result is
\begin{equation}
   \hat D_2 = {2 A\over(D-4)(D-3)}\,\big[2(y+1)\big]^{D/2-2} \,.
\end{equation}
Obviously the third diagram, $D_3$, gives the same contribution.

Next we set $D=4+2\epsilon$ and expand in $\epsilon$, using some of the
integrals collected in Appendix~\ref{app:2}. We find
\begin{equation}
   \sum_{i=1}^3 \hat D_i = A\,\bigg\{
    {1\over\epsilon}\,\Big[2-y\,r(y)\Big] + 2\Big[1-y\,r(y)\Big]
    \ln\big[2(y+1)\big] + 2 y h(y) - 4 + {\cal{O}}(\epsilon) \bigg\}
    \,.
\end{equation}
The functions $r(y)$ and $h(y)$ are given by
\begin{eqnarray}\label{rdef}
   r(y) &=& {\ln(y_+)\over\sqrt{y^2-1}} \,, \nonumber\\
   h(y) &=& {1\over\sqrt{y^2-1}}\,\Big[ L_2(1-y_-^2) - L_2(1-y_-) \Big]
    + {3\over 4}\sqrt{y^2-1}\,r^2(y) \,,
\end{eqnarray}
where $y_\pm=y\pm\sqrt{y^2-1}$, and $L_2(x)$ is the dilogarithm (see
Appendix~\ref{app:2}). They satisfy $r(1)=h(1)=1$.

\subsection{\bf Gluons attached to the light quark line}

Next consider the self-energy contribution for the light quark shown
in diagram $D_4$ in Fig.~\ref{fig:1}. It gives
\begin{eqnarray}
   D_4 &=& -4 (D-2) N_c C_F\,g_s^2\,{\rm Tr}\big\{\,
    \overline{\cal{P}}'\Gamma\,{\cal{P}}\,\gamma^\alpha\gamma^\beta
    \gamma^\gamma\big\}
    \nonumber\\
   &&\times \int\!{\rm d}\tilde s\,{\rm d}\tilde t\,
    {s_\alpha\,t_\beta\,s_\gamma\over(\omega+2v\cdot s)
    (\omega'+2v'\cdot s) \big(s^2\big)^2\,t^2\,(s-t)^2} \,.
\end{eqnarray}
After performing the integral over $t$, the remaining one-loop integral
can be readily evaluated using the master equation (\ref{master1}).
After Borel transformation, one finds the simple result
\begin{equation}
   \hat D_4 = - {A\over D-4} = - {A\over 2\epsilon} \,.
\end{equation}

\subsection{\bf Gluons attached to both heavy and light quark lines}

Let us now turn to the most cumbersome part of the calculation, namely
the loop correction of the heavy-light vertices. The contribution of
the diagram $D_5$ in Fig.~\ref{fig:1} is
\begin{eqnarray}
   D_5 &=& 8 N_c C_F\,g_s^2\,{\rm Tr}\big\{\,\overline{\cal{P}}'
    \Gamma\,{\cal{P}}\,\gamma^\alpha\rlap/v\gamma^\beta\big\}
    \nonumber\\
   &&\times \int\!{\rm d}\tilde s\,{\rm d}\tilde t\,
    {t_\alpha\,s_\beta\over(\omega+2v\cdot s) (\omega+2v\cdot t)
    (\omega'+2v'\cdot s)\,s^2\,t^2\,(s-t)^2} \,.
\end{eqnarray}
It is convenient to split the calculation into three parts by use of
the trace identity
\begin{equation}
   {\rm Tr}\big\{\,\overline{\cal{P}}'\Gamma\,{\cal{P}}\,
   \gamma^\alpha\rlap/v\gamma^\beta\big\} =
    {\rm Tr}\Big\{\,\overline{\cal{P}}'\Gamma\,{\cal{P}}
    \Big(g^{\alpha\beta} + \case{1}/{2}\,[\gamma^\alpha,\gamma^\beta]
    + 2 v^\alpha\gamma^\beta\Big) \Big\} \,.
\end{equation}
Let us denote the corresponding contributions by $D_5^{(i)}$ and
discuss them in turn.

\subsubsection{Calculation of $D_5^{(1)}$}

By rewriting its numerator, the integral appearing in $D_5^{(1)}$ can
be further decomposed into three parts:
\begin{equation}
   \int\!{\rm d}\tilde s\,{\rm d}\tilde t\,
   {s^2+t^2-(s-t)^2\over(\omega+2v\cdot s) (\omega+2v\cdot t)
   (\omega'+2v'\cdot s)\,s^2\,t^2\,(s-t)^2} \equiv I_1 + I_2 + I_3 \,.
\end{equation}
The integral $I_1$ can be evaluated by introducing a new variable
$s'=s+t$ and using the master equation (\ref{master1}). We find
\begin{eqnarray}
   I_1 &=& {i(-1)^{D-3}\over(4\pi)^{D/2}}\,\Gamma(4-D)\,
    \Gamma\big(\case{D}/{2}-1\big)
    \int\limits_0^\infty\!{\rm d}u\,{1\over V^2} \nonumber\\
   &&\times \int\!{\rm d}\tilde t\,{1\over(\omega+2v\cdot t)
    \big(\Omega/V+2\hat V\cdot t\big)^{4-D}\,t^2} \,,
\end{eqnarray}
where $V=(1+2yu+u^2)^{1/2}$, $\Omega=\omega+u\omega'$, and
$\hat V_\alpha=(v+u v')_\alpha/V$ is a unit vector. The integral over
$t$ can again be performed using the master equation, resulting in a
double parameter integral. The result simplifies upon Borel
transformation. We find
\begin{equation}\label{Cdef}
   \hat I_1 = {2 C\over(D-2)}\,{G\big(0,0,\case{D}/{2}-1;y\big)
   \over\big[2(y+1)\big]^{D/2-1}} ~;~~
   C \equiv {(2T)^{2D-5}\over(4\pi)^D}\,
   \Gamma\big(\case{D}/{2}\big)\,\Gamma\big(\case{D}/{2}-1\big) \,.
\end{equation}

The calculation of the remaining integrals is straightforward. For
$I_2$ we introduce $t'=t+s$ and make repeated use of the master
equations. $I_3$ factorizes into the product of one-loop integrals. To
evaluate its Borel transform it is convenient to combine denominators
by a Feynman parameter. This gives
\begin{eqnarray}\label{I23res}
   \hat I_2 &=& - {2 C\over(D-3)(D-2)}\,
    {1\over\big[2(y+1)\big]^{D/2-1}} \,, \nonumber\\
   \hat I_3 &=& -{2 C\over(D-2)}\,G\big(2-D,0,\case{D}/{2}-1;y\big) \,.
\end{eqnarray}
The calculation of the parameter integral in $\hat I_3$ is discussed in
Appendix~\ref{app:2}.

\subsubsection{Calculation of $D_5^{(2)}$}

For the two-loop integral appearing in $D_5^{(2)}$ we use the integral
representations discussed in Appendix~\ref{app:1}. We find that
\begin{eqnarray}\label{I4def}
   &&\case{1}/{2}\,{\rm Tr}\big\{\,\overline{\cal{P}}'\Gamma\,
    {\cal{P}}\,[\gamma^\alpha,\gamma^\beta]\,\big\}
    \int\!{\rm d}\tilde s\,{\rm d}\tilde t\,
    {t_\alpha\,s_\beta\over(\omega+2v\cdot s) (\omega+2v\cdot t)
    (\omega'+2v'\cdot s)\,s^2\,t^2\,(s-t)^2} \nonumber\\
   &&= (y-1)\,{\rm Tr}\big\{\,\overline{\cal{P}}'\Gamma\,
    {\cal{P}}\,\big\}\,I_4 \,,
\end{eqnarray}
where after Borel transformation
\begin{eqnarray}\label{I4res}
   \hat I_4 = {C\over D-3} &\Bigg\{& G\big(3-D,0,\case{D}/{2};y\big)
    - {G\big(1,0,\case{D}/{2};y\big)\over\big[2(y+1)\big]^{D/2-1}}
    \nonumber\\
   &&- {\Gamma(D-1)\over\Gamma\big(\case{D}/{2}\big)\,
        \Gamma\big(\case{D}/{2}-1\big)}
    \int\limits_1^\infty\!{\rm d}u_1 \int\limits_1^\infty\!{\rm d}u_2
    \,{(u_1 u_2-1)^{D/2-2}\over\big[u_1+2(y+1)(u_2-1)\big]^{D-1}}
    \Bigg\} \,.
\end{eqnarray}
The double integral becomes trivial in the limit $D\to 4$. The
evaluation of the first parameter integral is outlined in
Appendix~\ref{app:2}.

Remarkable cancellations take place when one adds up the contributions
from $I_1$ to $I_4$. We find the simple result
\begin{equation}
   \hat D_5^{(1)} + \hat D_5^{(2)} = A\,\bigg\{ - {1\over 2\epsilon}
   + {\cal{O}}(\epsilon) \bigg\} \,.
\end{equation}

\subsubsection{Calculation of $D_5^{(3)}$}

This part of the amplitude involves the hardest integral:
\begin{equation}
   I_\beta = \int\!{\rm d}\tilde s\,{\rm d}\tilde t\,
   {2v\cdot t\,s_\beta\over(\omega+2v\cdot s) (\omega+2v\cdot t)
   (\omega'+2v'\cdot s)\,s^2\,t^2\,(s-t)^2} \,.
\end{equation}
It can be simplified by using the method of integration by parts, which
allows one to reduce a given loop integral to a series of simpler
integrals \cite{IbP1,IbP2,IbP3}. In this case, we can relate $I_\beta$
to a sum of four integrals involving five (instead of six) types of
propagators. To this end, we evaluate the identity
\begin{equation}
   \int\!{\rm d}\tilde s\,{\rm d}\tilde t\,
   {\partial\over\partial t_\alpha}\,
   {2v\cdot t\,s_\beta\,(t-s)_\alpha\over(\omega+2v\cdot s)
   (\omega+2v\cdot t) (\omega'+2v'\cdot s)\,s^2\,t^2\,(s-t)^2} = 0
\end{equation}
to obtain
\begin{eqnarray}\label{Jdef}
   -(D-4)\,I_\beta &=& \int\!{\rm d}\tilde s\,{\rm d}\tilde t\,
    {s_\beta\over(\omega+2v\cdot s) (\omega'+2v'\cdot s)\,
    s^2\,t^2\,(s-t)^2} \nonumber\\
   &+& \int\!{\rm d}\tilde s\,{\rm d}\tilde t\,
    {2v\cdot t\,s_\beta\over(\omega+2v\cdot s) (\omega+2v\cdot t)
    (\omega'+2v'\cdot s) \big(t^2\big)^2\,(s-t)^2} \nonumber\\
   &-& \int\!{\rm d}\tilde s\,{\rm d}\tilde t\,
    {2v\cdot t\,s_\beta\over(\omega+2v\cdot s) (\omega+2v\cdot t)
    (\omega'+2v'\cdot s) \big(t^2\big)^2\,s^2} \nonumber\\
   &-& \int\!{\rm d}\tilde s\,{\rm d}\tilde t\,
    {\omega\,s_\beta\over(\omega+2v\cdot t)^2 (\omega'+2v'\cdot s)\,
    s^2\,t^2\,(s-t)^2} \nonumber\\
   &=& J_\beta^{(1)} + J_\beta^{(2)} + J_\beta^{(3)}
    + J_\beta^{(4)} \,.
\end{eqnarray}
The first three integrals can be calculated along the lines discussed
above. After Borel transformation, we find
\begin{eqnarray}
   \hat J_\beta^{(1)} &=& {4 C\over(D-4)(D-2)}\,
    {(v+v')_\beta\over\big[2(y+1)\big]^{D-2}} \,, \nonumber\\
   \hat J_\beta^{(2)} &=& -{C\over\big[2(y+1)\big]^{D/2-2}}
    \bigg\{ \bigg[ (v+v')_\beta - {2v_\beta\over D-2} \bigg]\,
    {G\big(0,0,\case{D}/{2}-1;y\big)\over 2(y+1)} \nonumber\\
   &&\phantom{ -{C\over\big[2(y+1)\big]^{D/2-2}}\bigg\{ }
    + G\big(0,1,\case{D}/{2};y\big)\,v_\beta
    + G\big(0,0,\case{D}/{2};y\big)\,v'_\beta \bigg\} \,,
    \nonumber\\
   \hat J_\beta^{(3)} &=& 2 C\,
    \Big[ G\big(3-D,1,\case{D}/{2}\big)\,v_\beta
    + G\big(3-D,0,\case{D}/{2}\big)\,v'_\beta \Big] \,,
\end{eqnarray}
with $C$ as defined in (\ref{Cdef}). Notice that because of the factor
$(D-4)$ on the left-hand side of (\ref{Jdef}) these expressions have to
be evaluated up to first order in $\epsilon$.

The evaluation  of $J_\beta^{(4)}$ is more involved. We have used the
method of differential equations to calculate this integral \cite{DEq}.
This interesting technique will be discussed below. For the moment we
just present the result:
\begin{eqnarray}\label{J4}
   \hat J_\beta^{(4)} &=& -{2 C v'_\beta\over D-2} \Bigg\{
    {1\over\big[2(y+1)\big]^{D/2-1}} -
    (D-4)\,G\big(0,D-4,\case{D}/{2}-1;y\big) \Bigg\} \nonumber\\
   &&- 2 C \int\limits_0^1\!{\rm d}u\,\Big(1-u^{D-4}\Big)\,
    {(u v+v')_\beta\over\big(1+2yu+u^2\big)^{D/2}} \,.
\end{eqnarray}
We have not written the last integral in terms of $G$-functions in
order to show explicitly that it is of order $(D-4)$.

Next we set $D=4+2\epsilon$ and expand the above expressions, keeping
terms of order $\epsilon$. The integrals encountered are collected in
Appendix~\ref{app:2}. They yield rather nontrivial functions of $y$.
However, again remarkable cancellations appear if one sums the various
contributions to the right-hand side of (\ref{Jdef}). Not only do the
poles in $1/\epsilon$ contained in $J_\beta^{(1)}$ and $J_\beta^{(3)}$
cancel, but also most of the $y$-dependent terms. Our final result is,
in fact, very simple. It reads
\begin{eqnarray}
   -(D-4)\,\hat I_\beta\,{\rm Tr}\big\{\,\overline{\cal{P}}'\Gamma\,
   {\cal{P}}\,\gamma^\beta\big\} &=& {C\over\big[2(y+1)\big]^{D-2}}\,
    {\rm Tr}\big\{\,\overline{\cal{P}}'\Gamma\,{\cal{P}}\,\big\}
    \\
   &&\times\bigg\{ 4 - 2\epsilon \bigg[ 2 + {2\pi^2\over 3}
    + (y^2-1)\,r^2(y) \bigg] + {\cal{O}}(\epsilon^2) \bigg\} \,.
    \nonumber
\end{eqnarray}
After Borel transformation, the diagram $D_6$ gives the same
contribution as $D_5$. Hence
\begin{equation}
   \hat D_5 + \hat D_6 = A\,\bigg\{ {1\over\epsilon} - 2
   - {2\pi^2\over 3} - (y^2-1)\,r^2(y) + {\cal{O}}(\epsilon) \bigg\} \,.
\end{equation}

\subsection{\bf Calculation of $J_\beta^{(4)}$ using a differential
equation}

In this paragraph we illustrate the application of differential
equations to the analysis of multi-loop diagrams. Such techniques were
introduced in Ref.~\cite{DEq} to evaluate integrals with massive
propagators. They are readily adaptable to HQET. The idea is to derive
a differential equation for a particular loop integral whose
inhomogeneous term can be calculated in terms of simpler integrals. The
original loop integral is then obtained from the solution of the
differential equation.

We start by rewriting $J_\beta^{(4)} = x \case{\partial}/{\partial x}
J_\beta(x)\vert_{x=1}$, where
\begin{equation}\label{Jx}
   J_\beta(x) = \int\!{\rm d}\tilde s\,{\rm d}\tilde t\,
   {s_\beta\over(x\omega+2v\cdot t) (\omega'+2v'\cdot s)\,
   s^2\,t^2\,(s-t)^2} \,.
\end{equation}
To derive a differential equation for $J_\beta(x)$, we use again the
method of integration by parts. Starting from the identity
\begin{equation}
   \int\!{\rm d}\tilde s\,{\rm d}\tilde t\,
   {\partial\over\partial t_\alpha}\,{s_\beta\,t_\alpha\over
   (x\omega+2v\cdot t) (\omega'+2v'\cdot s)\,s^2\,t^2\,(s-t)^2} = 0
\end{equation}
we find
\begin{equation}
   \bigg[ (D-4) - x {\partial\over\partial x} \bigg] J_\beta(x)
   = f_\beta(x) \,,
\end{equation}
where
\begin{eqnarray}
   f_\beta(x) &=& \int\!{\rm d}\tilde s\,{\rm d}\tilde t\,
    {s_\beta\over(x\omega+2v\cdot t) (\omega'+2v'\cdot s)\,s^2\,
    \big[(s-t)^2\big]^2} \nonumber\\
   &-& \int\!{\rm d}\tilde s\,{\rm d}\tilde t\,
    {s_\beta\over(x\omega+2v\cdot t) (\omega'+2v'\cdot s)\,t^2\,
    \big[(s-t)^2\big]^2}
\end{eqnarray}
does indeed only contain simpler integrals. The general solution of the
differential equation is
\begin{equation}\label{solution}
   J_\beta(x) = x^{D-4} \bigg\{ \int\limits_x^\infty\!{\rm d}z\,z^{3-D}
   f_\beta(z) + k_\beta \bigg\} \,,
\end{equation}
with $k_\beta$ being independent of $x$. We are interested in the Borel
transform of this equation. A straightforward calculation gives
\begin{eqnarray}
   z^{3-D} \hat f_\beta(z) = {(2T)^{2D-5}\over(4\pi)^D}
   \Gamma^2\big(\case{D}/{2}-1\big) &\Bigg\{&
    {v'_\beta\over\big(1+2yz+z^2\big)^{D/2-1}} \\
   &&- {D-2\over D-4}\,\Big(1-z^{4-D}\Big)\,
    {z^{D-3}\,(v+zv')_\beta\over\big(1+2yz+z^2\big)^{D/2}} \Bigg\} \,.
    \nonumber
\end{eqnarray}
In order to determine the constant of integration $k_\beta$ we consider
the limit $x\to\infty$ in (\ref{Jx}), in which
\begin{eqnarray}
   \lim_{x\to\infty} x J_\beta(x) &=& {1\over\omega}
    \int\!{\rm d}\tilde s\,{\rm d}\tilde t\,
    {s_\beta\over(\omega'+2v'\cdot s)\,s^2\,t^2\,(s-t)^2} \nonumber\\
   &\stackrel{\rm B.T.}{\to}& {(2T)^{2D-5}\over(4\pi)^D}\,
    \Gamma\big(\case{D}/{2}-1\big)\,\Gamma\big(\case{D}/{2}-2\big)\,
    v'_\beta \,.
\end{eqnarray}
By evaluating (\ref{solution}) for $x\gg 1$, on the other hand, we
find (for $D>3$)
\begin{equation}
   \lim_{x\to\infty} x \hat J_\beta(x) = {(2T)^{2D-5}\over(4\pi)^D}\,
   \Gamma\big(\case{D}/{2}-1\big)\,\Gamma\big(\case{D}/{2}-2\big)\,
   v'_\beta + \hat k_\beta \lim_{x\to\infty} x^{D-3} \,.
\end{equation}
Hence $\hat k_\beta=0$ follows. From the solution of the differential
equation we now obtain
\begin{equation}
   \hat J_\beta^{(4)} = x {\partial\over\partial x} \hat J_\beta(x)
   \bigg\vert_{x=1} = -\hat f_\beta(1)
   + (D-4) \int\limits_1^\infty\!{\rm d}z\,z^{3-D} \hat f_\beta(z) \,.
\end{equation}
Substituting here $z=1/u$ leads to (\ref{J4}).

\subsection{\bf Summary and renormalization}

We are now in a position to sum up the various two-loop corrections
that contribute at order $\alpha_s$ to the Borel-transformed
correlator. We find
\begin{eqnarray}
   \hat\Xi_1 \equiv \sum_{i=1}^6 \hat D_i = A &\bigg\{&
    {3\over 2}\bigg({1\over\epsilon}-{4\pi^2\over 9}-{8\over 3}\bigg)
    - \Big[ y\,r(y)-1 \Big] \bigg({1\over\epsilon} +
    2\ln\big[2(y+1)\big]\bigg) \nonumber\\
   &&+ 2 \Big[ y\,h(y)-1 \Big] - (y^2-1)\,r^2(y)
    + {\cal{O}}(\epsilon) \bigg\} \,.
\end{eqnarray}
Next we expand $A$ from (\ref{Adef}) around $D=4$, keeping terms of
order $\epsilon$, and relate $\hat\Xi_1$ to the lowest-order correlator
\begin{equation}
   \hat\Xi_0 = {3 T^3\over 4\pi^2} \bigg({2\over y+1}\bigg)^2\,
   {\rm Tr}\big\{\,\overline{\cal{P}}'\Gamma\,{\cal{P}}\,\big\} \,.
\end{equation}
This yields
\begin{eqnarray}\label{Xi1}
   \hat\Xi_1 = {\alpha_s\over\pi}\,\hat\Xi_0
   &\Bigg\{& -{1\over\hat\epsilon}
    + \bigg(2\ln{\mu\over T} + 2\gamma_E - 3\bigg)
    + {4\pi^2\over 9} + {17\over 3} \nonumber\\
   &&- {\gamma(y)\over 2} \bigg[ -{1\over\hat\epsilon}
    + \bigg(2\ln{\mu\over T} + 2\gamma_E - 3\bigg) \bigg]
    + c_{\rm pert}(y) + {\cal{O}}(\epsilon) \Bigg\} \,,
\end{eqnarray}
where
\begin{equation}
   {1\over\hat\epsilon} = {1\over\epsilon} + \gamma_E
   - \ln{4\pi\over\mu^2} \,.
\end{equation}
We have introduced the functions
\begin{eqnarray}\label{cpert}
   \gamma(y) &=& {4\over 3} \Big[ y\,r(y)-1 \Big] \,, \\
   c_{\rm pert}(y) &=& {\gamma(y)\over 2}
    \bigg[ 4\ln 2 - 3 + \ln{y+1\over 2} \bigg]
    - {4\over 3} \Big[ y\,h(y)-1 \Big] + \ln{y+1\over 2}
    + {2\over 3}\,(y^2-1)\,r^2(y) \nonumber\\
   &=& \bigg({16\over 9}\ln 2 - {49\over 54}\bigg)(y-1)
    - \bigg({8\over 15}\ln 2 - {197\over 600}\bigg)(y-1)^2
    + \ldots \,, \nonumber
\end{eqnarray}
both of which vanish at $y=1$. The first two terms in the expansion of
$c_{\rm pert}(y)$ around zero recoil were previously calculated in
Ref.~\cite{Patricia}, and we confirm the result obtained there.

The $1/\hat\epsilon$ poles in (\ref{Xi1}) cancel upon renormalization
of the heavy-light and heavy-heavy currents in (\ref{correl}). In the
$\overline{\rm MS}$ subtraction scheme, the corresponding
renormalization factors are \cite{Volo,Poli,Falk}
\begin{equation}
   {\cal{Z}}_{\rm hl} = 1 - {\alpha_s\over 2\pi\hat\epsilon} ~,~~
   {\cal{Z}}_{\rm hh} = 1 + {\alpha_s\over 2\pi\hat\epsilon}\,
   \gamma(y) \,.
\end{equation}
That means that our two-loop calculation reproduces correctly  the
known running of the hadronic form factors $F(\mu)$ and $\xi(y,\mu)$.
By comparing (\ref{Xi1}) with (\ref{wdep}) we can now write our final
result for the renormalized correlator in form of a dispersion integral
and introduce the continuum threshold $\omega_0$ to obtain
\begin{eqnarray}\label{Kperty}
   K_{\rm pert}(T,\omega_0,y) &=& {3\over 8\pi^2}
    \bigg({2\over y+1}\bigg)^2 \int\limits_0^{\omega_0}\!
    {\rm d}\omega_+\,\omega_+^2\,e^{-\omega_+/T} \nonumber\\
   &&\times\bigg\{ 1 + {\alpha_s\over\pi} \bigg[
    2\ln{\mu\over\omega_+} + {4\pi^2\over 9} + {17\over 3}
    - \gamma(y)\,\ln{\mu\over\omega_+} + c_{\rm pert}(y) \bigg]
    \bigg\} \,.
\end{eqnarray}
This is the exact expression for the perturbative part of the
correlator at order $\alpha_s$. It is now seen that at zero recoil one
indeed recovers (\ref{Kpert1}).

\section{Renormalization-group improvement\\
         and numerical analysis}
\label{sec:4}

The theoretical expression for the correlator depends on the
subtraction scale $\mu$, indicating a scheme-dependence associated with
the subtraction of the $1/\hat\epsilon$ poles. This just reflects that
the hadronic parameters $F(\mu)$ and $\xi(y,\mu)$, which are defined in
terms of matrix elements of currents in the effective theory, are
scheme-dependent quantities. At next-to-leading order in
renormalization-group improved perturbation theory one can define
renormalized, scheme-independent form factors by \cite{QCD}
\begin{eqnarray}
   F_{\rm ren} &=& \Big[\alpha_s(\mu)\Big]^{2/9}
    \bigg\{1 - {\alpha_s(\mu)\over\pi}\,
    \Big[ Z_{\rm hl} + \delta_{\rm hl} \Big] \bigg\}\,F(\mu) \,,
    \nonumber\\
   \xi_{\rm ren}(y) &=& \Big[\alpha_s(\mu)\Big]^{-a_L(y)}
    \bigg\{1 - {\alpha_s(\mu)\over\pi}\,
    \Big[ Z_{\rm hh}(y) + \delta_{\rm hh}(y) \Big] \bigg\}\,
    \xi(y,\mu) \,,
\end{eqnarray}
where $a_L(y)=\case{2}/{9}\,\gamma(y)$ \cite{Falk}, and we have used
that the number of light quark flavors in the effective theory is
$n_f=3$. The next-to-leading logarithmic corrections consist of two
parts. The coefficients $Z$ are renormalization-group invariant
quantities. For $n_f=3$, they are given by \cite{QCD,Broad,JiMu,RaKo}
\begin{eqnarray}
   Z_{\rm hl} &=& -{185\over 324} - {7\pi^2\over 243} \,, \nonumber\\
   \\
   Z_{\rm hh}(y) &=& {\gamma(y)\over 3}
    \bigg\{ {53\over 54} - {\pi^2\over 12} - y\,r(y)
    + {y\over\sqrt{y^2-1}}\,\Big[ L_2(1-y_-^2) + \ln^2(y_-) \Big]
    \bigg\} \nonumber\\
   &&- {8\over 9} \int\limits_0^\theta\!{\rm d}\psi
    \Big[ \psi\coth\psi - 1 \Big] \bigg\{ \psi\coth^2\!\theta +
    {\sinh\theta\,\cosh\theta\over\sinh^2\!\theta - \sinh^2\!\psi}\,
    \ln{\sinh\theta\over\sinh\psi} \bigg\} \nonumber\\
   &=& \bigg({752\over 729} - {8\pi^2\over 81}\bigg)(y-1)
    - \bigg({368\over 1215} - {4\pi^2\over 135}\bigg)(y-1)^2
    + \ldots \,, \nonumber
\end{eqnarray}
where $y_-=y-\sqrt{y^2-1}$, and the hyperbolic angle $\theta$ is
defined by $y=\cosh\theta$. The coefficients $\delta$, on the other
hand, are scheme-dependent. They arise from matching of QCD onto the
effective theory and combine with the scheme-dependent terms in
(\ref{Kperty}) to give a renormalization-group invariant result. In the
$\overline{\rm MS}$ subtraction scheme one has \cite{QCD,JiMu}
\begin{equation}
   \delta_{\rm hl} = {2\over 3} ~,~~
   \delta_{\rm hh}(y) = 0 \,.
\end{equation}

After renormalization-group improvement, the sum rule for the
renormalized Isgur-Wise function takes the form
\begin{equation}\label{xirensum}
   \xi_{\rm ren}(y) = \Big[\alpha_s(T)\Big]^{-a_L(y)}\,\,
   {\widehat{K}(T,\omega_0,y)\over\widehat{K}(T,\omega_0,1)} \,,
\end{equation}
where we have summed the leading logarithms down to a characteristic
scale given by the Borel parameter $T$. The renormalization-group
invariant function $\widehat{K}$ is given by
\begin{eqnarray}\label{Khat}
   \widehat{K}(T,\omega_0,y) &=& {3 T^3\over 8\pi^2}
    \bigg({2\over y+1}\bigg)^2 \int\limits_0^{\omega_0/T}\!
    {\rm d}x\,x^2\,e^{-x} \nonumber\\
   &&\times \bigg\{ 1 + {\alpha_s(T)\over\pi}
    \bigg[ {4\pi^2\over 9} + {13\over 3} - 2 Z_{\rm hl}
    - \big[ 2-\gamma(y)\big] \ln x + c_{\rm pert}(y)
    - Z_{\rm hh}(y) \bigg] \bigg\} \nonumber\\
   &-& \langle\bar q q\rangle(T) \bigg\{ 1\!+\!{\alpha_s(T)\over\pi}
    \bigg[ {2\over 3} - 2 Z_{\rm hl} + \gamma(y)
    \Big[ {\rm Ei}\Big(\!-\!{\omega_0\over T}\Big)\!-\!\gamma_E \Big]
    + c_{\langle\bar q q\rangle}(y)\!-\!Z_{\rm hh}(y) \bigg] \bigg\}
    \nonumber\\
   &+& \bigg({y-1\over y+1}\bigg)
    {\langle\alpha_s GG\rangle\over 48\pi T}
    + {(2y+1)\over 3}\,{m_0^2\,\langle\bar q q\rangle\over 4 T^2} \,.
\end{eqnarray}
We have included the order-$\alpha_s$ corrections to the quark
condensate as calculated in Ref.~\cite{SR}.\footnote{The function
$c_{\langle\bar q q\rangle}(y)$ was called $\case{2}/{3}
c_{\overline{\rm MS}}(y)$ in Ref.~\cite{SR}. Note that we have
simplified the dilogarithms appearing in $h(y)$ as compared to this
reference.}
The function $c_{\langle\bar q q\rangle}(y)$ has a similar form as
$c_{\rm pert}(y)$. It reads
\begin{eqnarray}
   c_{\langle\bar q q\rangle}(y) &=& {\gamma(y)\over 2}
    \bigg[ 4\ln 2 + \ln{y+1\over 2} \bigg]
    - {4\over 3} \Big[ y\,h(y)-1 \Big] - {2\over 3}\,(y-1)\,r(y)
    \nonumber\\
   &=& \bigg({16\over 9}\ln 2 - {56\over 27}\bigg)(y-1)
    - \bigg({8\over 15}\ln 2 - {112\over 225}\bigg)(y-1)^2
    + \ldots \,.
\end{eqnarray}
A remark is in order concerning the appearance of $\gamma_E$ and the
exponential integral in (\ref{Khat}). The effective spectral density
for the quark condensate receives contributions proportional to
$\delta(\omega_+)$ and $1/\omega_+ + \delta(\omega_+)\ln(\omega_+)$.
The latter ones have to be regularized in the dispersion integral,
leading to
\begin{equation}
   \lim_{\eta\to +0} \int\limits_0^\infty\!{\rm d}\omega_+\,
   e^{-\omega_+/T}\,\bigg[ {1\over\omega_+ +\eta}
   + \delta(\omega_+-\eta) \ln{\omega_+\over\mu} \bigg]
   = \ln{T\over\mu} - \gamma_E \,.
\end{equation}
If one introduces a continuum threshold, one obtains an extra
contribution ${\rm Ei}(-\omega_0/T)$ from the second term, where
${\rm Ei}(-x) = -\int_x^\infty{{\rm d}t\over t} e^{-t}$ is the
exponential integral. This contribution is very small and has been
neglected in Ref.~\cite{SR}.

For the numerical analysis of the sum rule (\ref{xirensum}) we use the
vacuum condensates as given in (\ref{cond}), as well as
$\Lambda_{\overline{\rm MS}}=0.25$ GeV (for $n_f=3$) in the running
coupling $\alpha_s(T)$. In Fig.~\ref{fig:2}(a) we show the range of
predictions for the renormalized Isgur-Wise function obtained by
varying the continuum threshold over the range $2.0<\omega_0<2.6$ GeV,
and the Borel parameter inside the ``sum rule window'' $0.8<T<1.2$ GeV,
where the theoretical calculation is reliable. This window is
determined by requiring that the nonperturbative contributions to the
sum rule be less than 30\% of the perturbative ones ($T>0.8$ GeV), and
that the pole contribution account for at least 30\% of the
perturbative part of the correlator ($T<1.2$ GeV). The lower limit on
$T$ also ensures that $\alpha_s(T)$ is small enough to allow for a
perturbative expansion. The above range of values for $\omega_0$ was
obtained from the study of the correlator of two heavy-light currents.
As in previous analyses, we observe excellent stability of the sum
rule. The sensitivity of the Isgur-Wise function to different choices
of the continuum model is investigated in detail in
Refs.~\cite{SR,Blok}. We do not discuss this subject here, since we are
mainly interested in the effects of radiative corrections. We just note
that the theoretical uncertainty in the sum rule prediction is probably
larger than indicated by the width of the band in Fig.~\ref{fig:2}(a).

To study the importance of radiative corrections we first have to
relate $\xi_{\rm ren}(y)$ to a more ``physical'' form factor, which
includes the logarithmic dependence on the heavy quark masses.
Otherwise it is not possible to consider the limit $\alpha_s\to 0$. For
simplicity, we work with a single scale $\bar m$ and define
\begin{equation}\label{xiphys}
   \xi_{\rm phys}(\bar m,y) = \Big[\alpha_s(\bar m)\Big]^{a_L(y)}
    \bigg\{1 + {\alpha_s(\bar m)\over\pi}\,Z_{\rm hh}(y) \bigg\}\,
    \xi_{\rm ren}(y) \,.
\end{equation}
We use $\bar m\simeq 2.3$ GeV as a characteristic scale for $b\to c$
transitions \cite{QCD}. Fig.~\ref{fig:2}(b) shows the next-to-leading
order result for this form factor in comparison with the ``bare''
Isgur-Wise function computed by neglecting radiative corrections. We
also show the leading logarithmic approximation to $\xi_{\rm phys}
(\bar m,y)$, which is obtained by ignoring terms of order $\alpha_s$ in
(\ref{Khat}) and keeping only the first factor in (\ref{xiphys}). It is
apparent from this figure that the radiative corrections to the
Isgur-Wise function are well under control. The large $y$-independent
corrections, which enhance the sum rule prediction for the decay
constant $F$ by 50\% \cite{BGSR,SR,BBBD}, cancel out in the ratio
(\ref{xirensum}). The remaining recoil-dependent radiative corrections
are small. At large recoil, they tend to decrease the form factor by
$5-10\%$. Part of this effect comes from leading logarithms and is
associated with the velocity-dependent anomalous dimension $a_L(y)$ of
the heavy-heavy current in the effective theory.

\section{Conclusions}
\label{sec:5}

We have presented the complete QCD sum rule analysis of the Isgur-Wise
form factor $\xi(v\cdot v')$ at next-to-leading order in
renormalization-group improved perturbation theory. To this end, we
have derived the exact result for the two-loop corrections to the
triangle quark loop. Such a calculation, which was never done before
for a form factor of heavy mesons, becomes feasible by using the heavy
quark effective theory. We have developed some general techniques for
dealing with two-loop integrals with two different types of heavy quark
propagators. Using the method of integration by parts, complicated
integrals can be reduced to simpler ones in a recursive way. Integrals
which cannot be reduced any further can be evaluated by using
differential equations. We have applied this technique to the loop
corrections to the heavy-light vertices. We have also presented an
integral representation for two-loop integrals which is particularly
convenient for QCD sum rule calculations. These methods can be applied
to other sum rule calculations and will eventually lead to more
accurate predictions for heavy meson form factors than were available
before.

Our numerical analysis shows that, unlike in the case of meson decay
constants, radiative corrections to the Isgur-Wise function are small
and well under control. This is an important result which puts the sum
rule analysis of $\xi(v\cdot v')$ on a firm footing. The smallness of
the two-loop corrections in this particular case was not unexpected,
however, since the normalization of the Isgur-Wise function at zero
recoil prohibits any recoil-independent radiative effects. This does
not imply that such corrections are always negligible. In fact, some of
the universal functions appearing at order $1/m_Q$ in the heavy quark
expansion receive their leading contributions at order $\alpha_s$. Then
the two-loop perturbative contribution is important and cannot be
neglected for a reliable analysis \cite{chi2}.

\acknowledgements
It is a pleasure to thank Lance Dixon, Adam Falk, Patrick Huet, Yossi
Nir and Michael Peskin for helpful discussions. Financial support from
the BASF Aktiengesellschaft and from the German National Scholarship
Foundation is gratefully acknowledged. This work was also supported by
the Department of Energy under contract DE-AC03-76SF00515.

\newpage
\appendix{Loop integrals in HQET}
\label{app:1}

\subsection{\bf One-loop integrals}

We summarize some important equations for one-loop tensor integrals in
HQET. Integrals involving massless propagators only can be found,
{\it i.e.}, in Ref.~\cite{PaTa}. Integrals involving two types of heavy
quark propagators were considered in the second reference in
\cite{Sublea}. There the master equation
\begin{eqnarray}\label{master1}
   I_{\mu_1\ldots\mu_n}(\alpha,\beta,\gamma)
   &=& \int\!{\rm d}\tilde t\,{t_{\mu_1}\ldots t_{\mu_n}\over
    \big(-t^2\big)^\alpha \big(\omega+2v\cdot t\big)^\beta
    \big(\omega'+2v'\cdot t\big)^\gamma} \\
   &=& {i\over(4\pi)^{D/2}}\,I_n(\alpha,\beta,\gamma)\,
    \int\limits_0^\infty\!{\rm d}u\,
    {u^{\gamma-1}\over\big[\Omega(u)\big]^{\beta+\gamma}}\,
    \bigg[-{\Omega(u)\over V(u)}\bigg]^{D-2\alpha+n}\,
    K_{\mu_1\ldots\mu_n}(u) \,, \nonumber
\end{eqnarray}
was derived, where ${\rm d}\tilde t=(2\pi)^{-D}{\rm d}^D t$, and
\begin{eqnarray}
   I_n(\alpha,\beta,\gamma) &=&
    {\Gamma(2\alpha+\beta+\gamma-D-n)\,\Gamma(D/2-\alpha+n)\over
     \Gamma(\alpha)\,\Gamma(\beta)\,\Gamma(\gamma)} \,, \nonumber\\
   \Omega(u) &=& \omega + u\,\omega' \,, \nonumber\\
   V(u) &=& (1 + u^2 + 2u\,v\cdot v')^{1/2} \,. \nonumber
\end{eqnarray}
For $n=0,1,2$ the tensors $K_{\mu_1\ldots\mu_n}(u)$ are given by
\begin{eqnarray}
   K(u) &=& 1 \,, \nonumber\\
   K_\mu(u) &=& - \hat V_\mu(u) \,, \nonumber\\
   K_{\mu\nu}(u) &=& \hat V_\mu(u)\,\hat V_\nu(u)
    - {g^{\mu\nu}\over D-2\alpha+2} \,, \nonumber
\end{eqnarray}
with $\hat V_\mu(u)=(v + u v')_\mu/V(u)$ being a unit vector. We note
that the master equation is valid for arbitrary values of $\alpha$,
$\beta$, and $\gamma$.

In the case of one heavy quark the master equation reduces to
\cite{Sublea,Broad}
\begin{eqnarray}\label{master2}
   I_{\mu_1\ldots\mu_n}(\alpha,\beta)
   &=& \int\!{\rm d}\tilde t\,t_{\mu_1}\ldots t_{\mu_n}\,
    \bigg({1\over-t^2}\bigg)^\alpha
    \bigg({\omega\over\omega+2v\cdot t}\bigg)^\beta \nonumber\\
   &=& {i\over(4\pi)^{D/2}}\,I_n(\alpha,\beta)\,
    \big(-\omega\big)^{D-2\alpha+n}\,K_{\mu_1\ldots\mu_n} \,,
\end{eqnarray}
where
\[
   I_n(\alpha,\beta) = {\Gamma(2\alpha+\beta-D-n)\,\Gamma(D/2-\alpha+n)
   \over\Gamma(\alpha)\,\Gamma(\beta)} \,.
\]
$K_{\mu_1\ldots\mu_n}$ is obtained from above by replacing
$\hat V_\mu(u)$ by $v_\mu$.

The integral representation (\ref{master1}) is particularly convenient
for a Borel transformation in $\omega$ and $\omega'$. Defining the
Borel operator by
\[
   {1\over\tau}\,\hat B_\tau^{(\omega)}
   = \!\!\lim_{\matrix{ n\to\infty \cr
                        -\omega\to\infty \cr}}\!\!
   {\omega^n\over\Gamma(n)}\,
   \bigg(-{{\rm d}\over{\rm d}\omega}\bigg)^n
   ~;~~ \tau = {-\omega\over n}~{\rm fixed} \,,
\]
we note that
\[
   \hat B_{\tau'}^{(\omega')} \hat B_\tau^{(\omega)}\,
   \big[-\Omega(u)\big]^{-a} = {\tau^{2-a}\over\Gamma(a)}\,
   \delta\Big(u-{\tau\over\tau'}\Big) \,.
\]

\subsection{\bf Two-loop integrals}

Let us now turn to two-loop integrals in HQET. The case with one heavy
quark has been discussed in detail in Ref.~\cite{Broad}. Here we
consider integrals with two species of heavy quark propagators. They
have the general form
\begin{eqnarray}\label{loopint}
   &&I_{\mu_1\ldots\mu_m}^{\nu_1\ldots\nu_n}
    (\alpha,\beta,\gamma,\delta;a,b,c) = \\
   &&\!\!\!\int\!{\rm d}\tilde s\,{\rm d}\tilde t\,
   {s_{\mu_1}\ldots s_{\mu_m}\,t^{\nu_1}\ldots t^{\nu_n}\over
   \big(\omega\!+\!2v\cdot s\big)^\alpha
   \big(\omega\!+\!2v\cdot t\big)^\beta
   \big(\omega'\!+\!2v'\cdot s\big)^\gamma
   \big(\omega'\!+\!2v'\cdot t\big)^\delta
   \big(\!-\!s^2\big)^a \big(\!-\!t^2\big)^b
   \big[\!-\!(s-t)^2\big]^c} \nonumber
\end{eqnarray}
We will derive a representation for this integral which is particularly
convenient for further analysis. The first step consists in performing
a Wick rotation of the loop momenta,
\[
   s ~\to~ (i s^0,\vec s\,) ~,~~ t ~\to~ (it^0,\vec t\,) \,,
\]
so that
\begin{eqnarray}
   -s^2 &~\to~& (s^0)^2 + \vec s\,^2 \equiv s_E^2 \,, \nonumber\\
   v\cdot s &~\to~& i v^0 s^0 - \vec v\cdot\vec s
    \equiv i v_E\!\cdot\! s_E \,, \nonumber
\end{eqnarray}
where $s_E=(s^0,\vec s\,)$ and $v_E=(v^0,i\vec v\,)$ are vectors in a
Euclidean space. Note that this definition of a Euclidean velocity
ensures that $v_E^2=1$ and $v_E\cdot v_E'=y$, where $y=v\cdot v'$
in Minkowski space. After the Wick rotation, we represent the massless
propagators as Fourier integrals in a $D$-dimensional Euclidean space
and use an exponential integral representation for the heavy quark
propagators:
\begin{eqnarray}
   {1\over\big(s_E^2\big)^a} &=&
    {\Gamma\big(\case{D}/{2}-a\big)\over\pi^{D/2}\,\Gamma(a)}
    \int\!{\rm d}^D x\,{e^{2is_E\cdot x}\over\big(x^2\big)^{D/2-a}}
    \,, \nonumber\\
   {1\over\big(\omega+2iv_E\!\cdot\! s_E\big)^\alpha} &=&
    {(-1)^\alpha\over\Gamma(\alpha)}
    \int\limits_0^\infty\!{\rm d}\lambda\,\lambda^\alpha\,
    e^{\lambda(\omega+2iv_E\cdot s_E)} \,. \qquad (\omega<0) \nonumber
\end{eqnarray}
The most general two-loop integral involves three $D$-dimensional
integrations over $x_i$ and four one-dimensional integrations over
$\lambda_i$. The advantage of these representations is that the
integrals over the loop momenta can immediately be performed and give
rise to two $D$-dimensional $\delta$-functions, which eliminate two of
the integrations over $x_i$. Furthermore, note that two of the
integrals over $\lambda_i$ become trivial upon Borel transformation,
since
\[
   \hat B_\tau^{(\omega)}\,e^{\lambda\omega}
   = \delta\big(\lambda-\tau^{-1}\big) \,.
\]
The tensor structure in the numerator in (\ref{loopint}) can be
generated by taking derivatives with respect to $x_i$. Recall that for
every timelike index there is a factor $i$ encountered during the Wick
rotation. For every spacelike index, on the other hand, one encounters
a factor $i$ when rotating back to Minkowski space. Together with a
factor $i^2$ from the loop integrations there is thus a factor of
$i^{2+m+n}$ to be taken into account.

Let us illustrate this technique for some of the integrals encountered
in Sec.~\ref{sec:3}. We start with the integral in (\ref{D1}). It
contains only two types of massless propagators, and consequently there
are no integrations over $x_i$ left after evaluation the
$\delta$-functions arising from the loop integrations. In the notation
of (\ref{loopint}) we find
\begin{eqnarray}
   I_\alpha(1,1,1,1;1,0,1) &=& {\Gamma\big(\case{D}/{2}\big)\,
    \Gamma\big(\case{D}/{2}-1\big)\over(4\pi)^D}
    \int\limits_0^\infty\!{\rm d}\lambda_1\,{\rm d}\lambda_2\,
    {\rm d}\lambda_3\,{\rm d}\lambda_4\,
    e^{(\lambda_1+\lambda_2)\omega+(\lambda_3+\lambda_4)\omega'}
    \nonumber\\
   &&\times {x_{1\alpha}\over\big(x_1^2\big)^{D/2}}\,
   {1\over\big(x_2^2\big)^{D/2-1}} \,, \nonumber
\end{eqnarray}
where $x_1=(\lambda_1+\lambda_2)v+(\lambda_3+\lambda_4)v'$, and
$x_2=\lambda_2 v+\lambda_4 v'$. After Borel transformation we obtain
\[
   \hat I_\alpha(1,1,1,1;1,0,1)
   = {C (v+v')_\alpha\over\big[2(y+1)\big]^{D/2}}
   \int\limits_0^1\!{\rm d}z_1\,{\rm d}z_2\,
   {1\over\big(z_1^2 + z_2^2 + 2yz_1z_2\big)^{D/2-1}}
\]
with $C$ as defined in (\ref{Cdef}). Substituting $z_2=u z_1$, an
integration by parts in $z_1$ yields
\[
   \int\limits_0^1\!{\rm d}z_1\,{\rm d}z_2\,
   {1\over\big(z_1^2 + z_2^2 + 2yz_1z_2\big)^{D/2-1}}
   = -{2\over(D-4)}\,G\big(0,0,\case{D}/{2}-1;y\big) \,,
\]
which leads to (\ref{D1res}). The integral $J_\beta^{(2)}$ in
(\ref{Jdef}) can be evaluated along the same lines. In this case there
are only three integrations over $\lambda_i$, and one is thus left with
a single parameter integral after Borel transformation.

A more complicated integral is that appearing in (\ref{I4def}).
Following the general procedure outlined above\, we derive
\begin{eqnarray}
   I_\beta^\alpha(1,1,1,0;1,1,1) &=&
    -{\Gamma^2\big(\case{D}/{2}\big)\,\Gamma\big(\case{D}/{2}-1\big)
    \over(4\pi)^{D/2}} \int\limits_0^\infty\!{\rm d}\lambda_1\,
    {\rm d}\lambda_2\,{\rm d}\lambda_3\,
    e^{(\lambda_1+\lambda_2)\omega+\lambda_3\omega'} \nonumber\\
   &&\times\int\!{\rm d}\tilde x_1\,
    {x_{1\beta}\over\big(x_1^2\big)^{D/2}}\,
    {x_2^\alpha\over\big(x_2^2\big)^{D/2}}\,
    {1\over\big(x_3^2\big)^{D/2-1}} \,, \nonumber
\end{eqnarray}
where $x_2=x_1+(\lambda_1+\lambda_2)v+\lambda_3 v'$, and
$x_3=x_1+\lambda_1 v+\lambda_3 v'$. The integral over $x_1$ has the
form of a Euclidean one-loop integral and can be performed in the
standard manner by introduction of two Feynman parameters $z_i$. One
then contracts the Lorentz indices with those in the trace in
(\ref{I4def}) to compute the integral $I_4$. The result is
\[
   I_4 = {\Gamma(D-1)\over(4\pi)^D}
   \int\limits_0^\infty\!{\rm d}\lambda_1\,{\rm d}\lambda_2\,
   {\rm d}\lambda_3\,\lambda_2\lambda_3\,
   e^{(\lambda_1+\lambda_2)\omega+\lambda_3\omega'}
   \int\!{\rm d}z_1\,{\rm d}z_2\,
   {z_1 (z_1^2 \bar z_1 z_2 \bar z_2)^{D/2-1}\over
    \big[ M^2(\lambda_i,z_i) \big]^{D-1}} \,,
\]
where $\bar z_i=1-z_i$, and
\[
   M^2(\lambda_i,z_i) = z_1 \big(z_2 p^2 + \bar z_2 q^2\big)
   - z_1^2 \big(z_2 p + \bar z_2 q\big)^2
\]
with $p=\lambda_2 v$ and $q=(\lambda_1+\lambda_2)v + \lambda_3 v'$.
After Borel transformation the integral can be cast into the following
form:
\[
   \hat I_4 = {(2T)^{2D-5}\over(4\pi)^D}\,\Gamma(D-1)
   \int\limits_0^1\!{\rm d}\lambda\,\lambda^{2-D}
   \int\limits_\lambda^\infty\!{\rm d}u_1
   \int\limits_{1/\lambda}^\infty\!{\rm d}u_2\,
   {(u_1 u_2-1)^{D/2-2}\over\big[ u_1+2(y+1)(u_2-1) \big]^{D-1}} \,.
\]
For $D<3$ one can use an integration by parts in $\lambda$ to obtain
(\ref{I4res}). By analytic continuation, this result can then be
evaluated around $D=4$.

\appendix{Parameter integrals}
\label{app:2}

We collect some useful formulae for the evaluation of parameter
integrals. We start with a remark on divergent integrals such as those
appearing in (\ref{I23res}) and (\ref{I4res}). Assuming first that $D$
is sufficiently small, one can use an integration by parts to rewrite
these in terms of integrals which have a well defined expansion around
$D=4$. For instance, for $D<4$ one can show that
\[
   (D-4)\,G\big(3-D,0,\case{D}/{2};y\big) = D\,
   \Big[ y\,G\big(4-D,0,\case{D}/{2}+1;y\big)
    + G\big(4-D,1,\case{D}/{2}+1;y\big) \Big] - 1 \,.
\]
Similarly, $G\big(2-D,0,\case{D}/{2}-1;y\big)$ can be related to
$G\big(3-D,0,\case{D}/{2};y\big)$ plus nonsingular terms for $D<3$. By
analytic continuation, one can then evaluate the resulting expressions
in the vicinity of $D=4$.

We now present a list of parameter integrals which are encountered when
one expands the results presented in Sec.~\ref{sec:3} around $D=4$.
When evaluating these integrals it is useful to introduce a hyperbolic
angle $\theta$ by $y=\cosh\theta$. Then
\[
   V^2(u) = 1+2yu+u^2 = (u+e^\theta)(u+e^{-\theta})
\]
factorizes. Setting $R_0=V^2(1)=2(y+1)$, we find:
\begin{eqnarray}
   F_1 &=& \int\limits_0^1\!{\rm d}u\,{1\over V^2(u)}
    = {r(y)\over 2} \,, \nonumber\\
   F_2 &=& R_0 \int\limits_0^1\!{\rm d}u\,{1\over\big[V^2(u)\big]^2}
    = {1-r(y)\over 2(y-1)} + 1 \,, \nonumber\\
   F_3 &=& R_0 \int\limits_0^1\!{\rm d}u\,{1+u\over\big[V^2(u)\big]^2}
    = -{y+1\over 2}\,r(y) - {\ln R_0 + l(y)\over 2} \,, \nonumber\\
   F_4 &=& R_0 \int\limits_0^1\!{\rm d}u\,
    {\ln(1-u)\over\big[V^2(u)\big]^2}
    = -{r(y)\over 4} - {1\over 4(y-1)}\,
    \Big\{ \big[r(y)+1\big]\,\ln R_0 - 2\,l(y) \Big\} \,, \nonumber\\
   F_5 &=& \int\limits_0^1\!{\rm d}u\,{\ln V^2(u)\over V^2(u)}
    = -{r(y)\over 2}\,\ln R_0 + h(y) \,, \nonumber\\
   F_6 &=& R_0 \int\limits_0^1\!{\rm d}u\,
    {\ln V^2(u)\over\big[V^2(u)\big]^2}
    = {1\over 2(y-1)}\,\Big\{ 2y - 2 h(y) + \big[r(y)-1\big]\,
    \big(1+\ln R_0\big) \Big\} \,, \nonumber\\
   F_7 &=& \int\limits_0^1\!{\rm d}u\,{\ln(1-u)\over(1-u)}
    \Bigg[ {R_0^2\over\big[V^2(u)\big]^2} - 1 \Bigg]
    = {\ln^2\!R_0\over 8} - {\pi^2\over 12} - {y^2-1\over 8}\,r^2(y)
    - {y+1\over 2}\,r(y) \nonumber\\
   &&- {y\,\ln R_0\over 2(y-1)} - {(2-y)(y+1)\over 4(y-1)}\,
    \Big[ r(y)\,\ln R_0 - 2\,l(y) \Big] \,, \nonumber\\
   F_8 &=& R_0^2 \int\limits_0^1\!{\rm d}u\,
    {\ln V^2(u)-\ln R_0\over (1-u)\,\big[V^2(u)\big]^2}
    = - {\pi^2\over 6} - {y^2-1\over 4}\,r^2(y)
    - {y+1\over 2}\,r(y)\,\ln R_0 \nonumber\\
   &&+ (y+1)\,\big(1-\ln R_0\big) - {1\over 2} + (y+1)\,
    \Big[ F_5 + F_6 - F_2\,\ln R_0 \Big] \,. \nonumber
\end{eqnarray}
The functions $r(y)$ and $h(y)$ have been defined in (\ref{rdef}). In
addition, we have introduced
\[
   l(y) = {1\over\sqrt{y^2-1}} \Big[ L_2(-y+\sqrt{y^2-1})
   - L_2(-y-\sqrt{y^2-1}) \Big] \,,
\]
which satisfies $l(1)=2\ln 2$. Here $L_2(x)=-\int_0^x{{\rm d}t\over t}
\ln(1-t)$ is the dilogarithm. The first derivatives of these functions
at $y=1$ are $r'(1)=-\case{1}/{3}$, $h'(1)=\case{1}/{18}$, and
$l'(1)=\case{1}/{6}-\case{2}/{3}\ln 2$. Finally, we note the following
useful identity ($n\ge 1$):
\[
   \int\limits_0^1\!{\rm d}u\,\ln^n(1-u)
   {D(y+u)\over\big[V^2(u)\big]^{D/2+1}}
   = n \int\limits_0^1\!{\rm d}u\,{\ln^{n-1}(1-u)\over(1-u)}\,
   \Big[ R_0^{-D/2} - \big[V^2(u)\big]^{-D/2} \Big] \,.
\]

\figure{\label{fig:1}
Two-loop diagrams contributing at order $\alpha_s$ to the perturbative
part of the sum rule for the Isgur-Wise form factor. Heavy quark
propagators are drawn as double lines, while the wavy line represents
the weak current.}

\figure{\label{fig:2}
(a) Sum rule prediction for the renormalized Isgur-Wise function
$\xi_{\rm ren}(y)$. The width of the band arises from variation of
$\omega_0$ and $T$ as specified in the text.
(b) The ``physical'' form factor $\xi_{\rm phys}(\bar m,y)$ computed in
next-to-leading order in renormalization-group improved perturbation
theory (solid), in leading logarithmic approximation (dashed), and
without including any QCD corrections (dotted). We use the central
values $\omega_0=2.3$ GeV and $T=1.0$ GeV.}


\begin{references}

\bibitem {Isgu}
N. Isgur and M.B. Wise, Phys.\ Lett.\ B {\bf 232}, 113 (1989);
{\bf 237}, 527 (1990).

\bibitem {Volo}
M.B. Voloshin and M.A. Shifman, Yad.\ Fiz.\ {\bf 45}, 463 (1987)
[Sov.\ J.\ Nucl.\ Phys.\ {\bf 45}, 292 (1987)]; {\bf 47}, 801 (1988)
[{\bf 47}, 511 (1988)].

\bibitem {Geor}
H. Georgi, Phys.\ Lett.\ B {\bf 240}, 447 (1990).

\bibitem {Mann}
T. Mannel, W. Roberts and Z. Ryzak, Nucl.\ Phys.\ {\bf B368}, 204
(1992).

\bibitem {Falk}
A.F. Falk, H. Georgi, B. Grinstein, and M.B. Wise, Nucl.\ Phys.\
{\bf B343}, 1 (1990);
A.F. Falk, B. Grinstein, and M.E. Luke, Nucl.\ Phys.\ {\bf B357}, 185
(1991).

\bibitem {Luke}
M.E. Luke, Phys.\ Lett.\ B {\bf 252}, 447 (1990).

\bibitem {AMM}
A.F. Falk, M. Neubert, and M.E. Luke, SLAC preprint SLAC--PUB--5771,
1992 (to appear in Nucl.\ Phys. {\bf B}).

\bibitem {SVZ}
M.A. Shifman, A.I. Vainshtein, and V.I. Zakharov, Nucl.\ Phys.\ {\bf
B147}, 385 (1979); {\bf B147}, 448 (1979).

\bibitem {Buch}
M. Neubert, V. Rieckert, B. Stech, and Q.P. Xu, in {\it Heavy Flavours},
edited by A.J. Buras and M. Lindner, Advanced Series on Directions in
High Energy Physics (World Scientific, Singapore, 1992).

\bibitem {Rady}
A.V. Radyushkin, Phys.\ Lett.\ B {\bf 271}, 218 (1991).

\bibitem {BGSR}
D.J. Broadhurst and A.G. Grozin, Phys.\ Lett.\ B {\bf 274}, 421 (1992).

\bibitem {SR}
M. Neubert, Phys.\ Rev.\ D {\bf 45}, 2451 (1992).

\bibitem {BBBD}
E. Bagan, P. Ball, V.M. Braun, and H.G. Dosch, Phys.\ Lett.\ B
{\bf 278}, 457 (1992).

\bibitem {Sublea}
M. Neubert, Phys.\ Rev.\ D {\bf 46}, 1076 (1992); {\bf 46}, 3914 (1992).

\bibitem {Blok}
B. Blok and M. Shifman, Santa Barbara preprint NSF--ITP--92--100, 1992.

\bibitem {Baier}
V.N. Baier and A.G. Grozin, Budker preprint BUDKERINP 92--80, 1992.

\bibitem {chi2} M. Neubert, Z. Ligeti and Y. Nir, SLAC preprint
SLAC--PUB--5915 (1992).

\bibitem {Patricia}  E. Bagan, P. Ball and P. Gosdzinsky, Heidelberg
preprint HD--THEP--92--40 (1992).

\bibitem {IbP1}
F.V. Tkachov, Phys.\ Lett.\ B {\bf 100}, 65 (1981);
K.G. Chetyrkin and F.V. Tkachov, Nucl.\ Phys.\ {\bf B192}, 159 (1981).

\bibitem {IbP2}
D.I. Kazakov, Phys.\ Lett.\ B {\bf 133}, 406 (1983).

\bibitem {IbP3}
N. Gray {\it et al.}, Z.\ Phys.\ C {\bf 48}, 673 (1990).

\bibitem {DEq}
A.V. Kotikov, Phys.\ Lett.\ B {\bf 254}, 158 (1991), {\bf 259}, 314
(1991).

\bibitem {Poli}
H.D. Politzer and M.B. Wise, Phys.\ Lett.\ B {\bf 206}, 681 (1988);
{\bf 208}, 504 (1988).

\bibitem {QCD}
M. Neubert, Phys.\ Rev.\ D {\bf 46}, 2212 (1992).

\bibitem {Broad}
D.J. Broadhurst and A.G. Grozin, Phys.\ Lett.\ B {\bf 267}, 105 (1991).

\bibitem {JiMu}
X. Ji and M.J. Musolf, Phys.\ Lett.\ B {\bf 257}, 409 (1991).

\bibitem {RaKo}
G.P. Korchemsky and A.V. Radyushkin, Nucl.\ Phys.\ {\bf B283}, 342
(1987); G.P. Korchemsky, Mod.\ Phys.\ Lett.\ A {\bf 4}, 1257 (1989).

\bibitem {PaTa}
P. Pascual and R. Tarrach, {\it QCD: Renormalization for the
Practitioner}, Lecture Notes in Physics No.\  194
(Springer, Berlin--Heidelberg--New York--Tokyo, 1984).

\end{references}
\end{document}